\documentclass[12pt]{article}
\usepackage{epsfig}
\usepackage{cite}

\newcommand{\mysection}{\setcounter{equation}{0}\section}

\def\beq{\begin{equation}}
\def\eeq{\end{equation}}
\def\beqa{\begin{eqnarray}}
\def\eeqa{\end{eqnarray}}
 
\newlength{\dinwidth} \newlength{\dinmargin}
\begin{document}

\begin{center}
{\Large \bf Resummation for $s$-channel single-top production}
\end{center}
\vspace{2mm}
\begin{center}
{\large Nikolaos Kidonakis}\\
\vspace{2mm}
{\it Department of Physics, Kennesaw State University,\\
Kennesaw, GA 30144, USA}
\end{center}
 
\begin{abstract}
I present higher-order results for $s$-channel single-top-quark production that are derived from soft-gluon resummation. I show that the soft-gluon corrections provide the dominant contribution to the QCD corrections, and that they approximate very well the exact results through NNLO. Furthermore, approximate N$^3$LO corrections are computed by using the expansion of the resummed cross section to third order in the strong coupling. The calculation treats the final-state $b$-quarks as massless but I also show how to extend the formalism to the case of massive $b$-quarks. The higher-order corrections are large at LHC energies and they improve the theoretical accuracy.
\end{abstract}
 
\mysection{Introduction}

Single top-quark production in proton collisions is an important process as it offers unique opportunities for understanding the electroweak properties of the top quark, including a direct measurement of the $V_{tb}$ CKM matrix element, and for additional insights into electroweak theory. Thus, increased precision in theoretical predictions for single-top cross sections is crucial in the exploration of this sector of the Standard Model as well as in the search for new physics.

Cross sections for single top-quark production in the $s$-channel have been measured at Tevatron \cite{D02009,CDF2009,CDF2010,D02011,D02013,CDF2014,CDF2014p,CDFD02014,CDF2016,CDFD02015} and LHC energies \cite{ATLAS7,ATLAS8a,ATLAS8b,CMS7and8,ATLASCMS,ATLAS13}, and the process has been studied theoretically for over two decades at fixed order \cite{NLO1,NLO2,NLO3,NLO4,NLO5,NLO6,NLO7,NLO8,NNLO} and using resummation methods \cite{NKnlltev,NKnlllhc,NKnnll,PPN,HQ13,SYY,NKdiff,NKrev,NK3l}. Numerically, the $s$-channel cross sections are the smallest among the three single-top processes at LHC energies, with the $t$-channel being the largest and the $tW$ channel being second largest.

The NLO QCD corrections for $s$-channel production were presented in \cite{NLO1,NLO2,NLO3,NLO4,NLO5,NLO6,NLO7,NLO8} and they are large, providing around 35\% increase of the cross section at LHC energies. More recently, the NNLO QCD corrections were calculated in Ref. \cite{NNLO} in the approximation where the corrections for the heavy-quark and the light-quark lines are treated seperately and color exchanges between them are neglected (and furthermore, the final-state $b$-quark mass is set to zero). These NNLO corrections are also considerable, providing around 7\% increase at LHC energies, and they give a reduced scale dependence relative to NLO. 

There are large contributions to these higher-order QCD corrections from soft-gluon emission. These contributions can be formally resummed in moment space, and the resummed cross section can then be used as a generator of higher-order predictions for the physical cross section \cite{NKnlltev,NKnlllhc,NKnnll,PPN,HQ13,NKdiff,NKrev,NK3l,NKaN3LO} without a need for a prescription. Detailed analytical results for these soft-gluon corrections in hard-scattering processes can be found through N$^3$LO in Ref. \cite{NKaN3LO}. The resummation of soft-gluon corrections follows from renormalization-group evolution and it is given in terms of soft anomalous dimensions, which are matrices in the space of color exchanges. The soft anomalous dimension matrix for this process has been calculated at one loop \cite{NKnlltev,NKnnll,NK3l}, two loops \cite{NKnnll,NK3l}, and partially at three loops \cite{NK3l}. The results in Refs. \cite{NKnlltev,NKnnll,NK3l} assume massless final-state $b$-quarks. In the next section, I review these results and also present new explicit analytical results for the soft anomalous dimension matrix through two loops for the case of massive final-state $b$-quarks.

In Section 3, I present results for the total cross section for single-top and single-antitop $s$-channel production at various LHC energies using the formalism of Refs. \cite{NKnlltev,NKnlllhc,NKnnll,NK3l,NKaN3LO} and single-particle-inclusive kinematics. It is shown that soft-gluon corrections dominate the QCD corrections through NNLO and, thus, provide very good approximations to the exact higher-order cross sections. This is similar to what is known about top-antitop production through NNLO \cite{NKtt2l} (see also \cite{PPN,HQ13,NKrev}). It is also in line with all other top-quark processes where soft-gluon corrections have been calculated in this formalism and compared to known full corrections through NLO, i.e. $tW$ \cite{NKnlltev,NKnlllhc,NKtW,NKtW2}, $t$-channel single-top \cite{NKnlltev,NKnlllhc,NKtch}, $tqH$ \cite{MFNK,tqH}, $tq\gamma$ \cite{tqgamma}, $tqZ$ \cite{tqZ}, $t{\bar t}\gamma$ \cite{ttgamma}, $t{\bar t}W$ \cite{ttW}, $t{\bar t}Z$ \cite{ttZ}, and $t{\bar t}H$ \cite{ttH} production, and even some processes not involving the top quark (see e.g. \cite{1PIH,Wgamma}). We stress that although all top-quark processes studied have confirmed the dominance of the soft-gluon corrections at NLO, this is only the second time that the comparison is made at NNLO, the first time being for the $t{\bar t}$ process. This is due to the lack of full NNLO calculations for the other processes (though a few comparisons to some partial NNLO results have also been made for $t{\bar t}W$ \cite{ttW} and $t{\bar t}H$ \cite{ttH}). Therefore, the explicit demonstration of the dominance of soft-gluon corrections at NNLO for $s$-channel production is noteworthy even if it is not surprising. 

Furthermore, third-order soft-gluon corrections are added to the NNLO results (which assume massless $b$-quarks in the final state) to derive approximate N$^3$LO (aN$^3$LO) cross sections that provide the best theoretical prediction for the cross section. Uncertainties from scale variation and from parton distributions are also calculated, and the theoretical predictions are compared to data from the LHC. Conclusions are presented in Section 4.

\mysection{Resummation for $s$-channel single-top processes}

In this section, we discuss soft-gluon resummation for single-top production in the $s$-channel. We begin with general considerations and continue with the specific cases of treating the final-state $b$-quark as either a massless or a massive particle.

For the single-top $s$-channel partonic reactions $q(p_1)+ {\bar q'} (p_2) \to t(p_3)+ {\bar b}(p_4)$, which involve a timelike virtual $W$ boson, the partonic kinematical variables are $s=(p_1+p_2)^2$, $t=(p_1-p_3)^2$, and $u=(p_2-p_3)^2$, with the top quark being the observed particle. We also define the threshold variable $s_4=s+t+u-m_t^2-m_b^2$, where we denote the top-quark mass by $m_t$ and the $b$-quark mass by $m_b$. The same notation and formalism applies to the case of $s$-channel single-antitop processes, $q(p_1)+ {\bar q'} (p_2) \to {\bar t}(p_3)+ b(p_4)$. We note that the choice of the top quark as the observed particle differs from that in Ref. \cite{NKnnll}. This different choice has a numerical effect and it better reproduces the exact NNLO corrections as we will describe in Section 3.

For the incoming partons, we may write their momenta, $p_i^{\mu}$, in terms of their four-velocities, $v_i^{\mu}$, as $p_1^{\mu}=\sqrt{s} \, v_1^{\mu}/2$ and $p_2^{\mu}=\sqrt{s} \, v_2^{\mu}/2$. For the top quark, we have $p_3^{\mu}=(s+m_t^2-m_b^2) \, v_3^{\mu}/(2\sqrt{s})$, while for the bottom quark, $p_4^{\mu}=(s+m_b^2-m_t^2) \, v_4^{\mu}/(2\sqrt{s})$. The eikonal rules used in the loop integrals for the resummation calculations involve these four-velocities.

Under Laplace transforms, the cross section can be factorized into functions that describe universal soft and collinear emission from the incoming partons, a hard function $H_{q{\bar q'} \to t{\bar b}}$ that describes the hard scattering, and a soft function $S_{q{\bar q'} \to t{\bar b}}$ for noncollinear soft-gluon emission \cite{NKnlltev,NKnnll,PPN,HQ13,NKrev}.

The resummed cross section follows from the evolution of these functions, and it is given in terms of the transform variable, $N$, by 
\beqa
{\hat{\sigma}}_{q{\bar q'} \to t{\bar b}}^{res}(N) &=&   
\exp\left[ \sum_{i=q,{\bar q'}} E_i(N_i)\right] \; 
\exp \left[\sum_{i=q,{\bar q'}} 2\int_{\mu_F}^{\sqrt{s}} \frac{d\mu}{\mu}\;
\gamma_{i/i}\left({\tilde N}_i,\alpha_s(\mu)\right)\right] \;
\exp\left[ E'(N)\right] 
\nonumber\\ && \hspace{-10mm} \times \,
{\rm tr} \left \{H_{q{\bar q'} \to t{\bar b}} \left(\alpha_s(\sqrt{s})\right) \;
{\bar P} \, \exp \left[\int_{\sqrt{s}}^{{\sqrt{s}}/{\tilde N}} \frac{d\mu}{\mu} \;
\Gamma^{\dagger}_{S \, q{\bar q'} \to t{\bar b}} \left(\alpha_s(\mu)\right)\right] \;
S_{q{\bar q'} \to t{\bar b}} \left(\alpha_s(\sqrt{s}/{\tilde N}) \right) \right. 
 \nonumber\\ && \quad \left.\times \,
P \, \exp \left[\int_{\sqrt{s}}^{{\sqrt{s}}/{\tilde N}} 
\frac{d\mu}{\mu}\; \Gamma_{S \, q{\bar q'} \to t{\bar b}}
\left(\alpha_s(\mu)\right)\right] \right\} \, .
\label{resHS}
\eeqa
In the above expression, $N_q=N(m_b^2-u)/s$, $N_{\bar q'}=N(m_b^2-t)/s$, ${\tilde N}=N e^{\gamma_E}$ with $\gamma_E$ the Euler constant, and $P$ (${\bar P}$) indicates path ordering in the same (reverse) sense as $\mu$.

The initial-state exponent in Eq. (\ref{resHS}) resums universal contributions from the incoming partons \cite{GS87,CT89} and can be written in the form
\beq
E_i(N_i)=
\int^1_0 dz \frac{z^{N_i-1}-1}{1-z}\;
\left \{\int_1^{(1-z)^2} \frac{d\lambda}{\lambda}
A_i\left(\alpha_s(\lambda s)\right)
+D_i\left[\alpha_s((1-z)^2 s)\right]\right\} \, ,
\label{Eexp}
\eeq
where $A_i$ is the lightlike cusp anomalous dimension, while in the second exponent, $\gamma_{i/i}$ is the moment-space anomalous dimension of the ${\overline {\rm MS}}$ density $\phi_{i/i}$. We use the same notation and expressions for the functions $A_i$, $D_i$, and $\gamma_{i/i}$ as in Refs. \cite{HQ13,NKrev,MFNK,tqH,1PIH,Wgamma}. We note that for next-to-next-to-leading-logarithm (NNLL) resummation accuracy, $A_i$ and $D_i$ are needed at two loops and $\gamma_{i/i}$ at one loop.

The final-state exponent, $E'(N)$, in Eq. (\ref{resHS}) is 0 for the case of massive $b$-quarks, while for the case of massless $b$-quarks it can be written as
\beq
E'(N)=
\int^1_0 dz \frac{z^{N-1}-1}{1-z}\;
\left \{\int^{1-z}_{(1-z)^2} \frac{d\lambda}{\lambda}
A_q \left(\alpha_s\left(\lambda s\right)\right)
+B_q \left[\alpha_s((1-z)s)\right]
+D_q \left[\alpha_s((1-z)^2 s)\right]\right\} \, ,
\label{Epexp}
\eeq
where, again, we use the same notation and expressions for the functions in the exponent as in Refs. \cite{HQ13,NKrev,MFNK,tqH,1PIH,Wgamma}. At NNLL accuracy, the function $B_q$ is needed at two loops.

The hard function, $H_{q{\bar q'} \to t{\bar b}}$, and the soft function, $S_{q{\bar q'} \to t{\bar b}}$, are $2 \times 2$ matrices in the space of color exchanges for this process \cite{NKnlltev,NKnnll}. Finally, $\Gamma_{S \, q{\bar q'} \to t{\bar b}}$ is the soft anomalous dimension which controls the evolution of the soft function, and it is also a $2 \times 2$ matrix for this process \cite{NKnlltev,NKnnll,NK3l}. For NNLL resummation, the hard and soft functions are needed at NLO, while the soft anomalous dimension is needed at two loops \cite{NKnnll} (see also \cite{FNPY1,FNPY2,ZLWZ}). At NNLL accuracy, while the NNLO soft-gluon corrections can be fully calculated, the lowest two logarithmic powers of the N$^3$LO corrections are only partially known. However, these partial contributions are expected to dominate the lowest powers which in any case are subleading overall (see the detailed discussion in Sec. III B of Ref. \cite{NKtW2}).

\subsection{Soft anomalous dimension for massless final-state $b$-quarks}

Next, we discuss and provide expressions for the soft anomalous dimension for $s$-channel single-top processes through two loops. We write the perturbative expansion as $\Gamma_{S \, q{\bar q'} \to t{\bar b}}=\sum_{n=1}^{\infty} (\alpha_s/\pi)^n \Gamma_S^{(n)}$.

We begin with the case where we treat the final-state $b$-quarks as massless.
We choose a color basis $c_1=\delta_{a_1 a_2} \delta_{a_3 a_4}$
and $c_2=T^c_{a_2 a_1} T^c_{a_3 a_4}$.
Then, the four matrix elements at one loop are given by \cite{NKnlltev,NKnnll,NK3l}
\beqa
\Gamma_{S\, 11}^{(1)}&=&C_F \left[\ln\left(\frac{s-m_t^2}{m_t\sqrt{s}}\right)
-\frac{1}{2}\right] \, ,
\nonumber \\
\Gamma_{S\, 12}^{(1)}&=&\frac{C_F}{2N_c} \ln\left(\frac{t(t-m_t^2)}{u(u-m_t^2)}\right) \, , \quad \quad
\Gamma_{S\, 21}^{(1)}=\ln\left(\frac{t(t-m_t^2)}{u(u-m_t^2)}\right) \, ,
\nonumber \\
\Gamma_{S\, 22}^{(1)}&=&\left(C_F-\frac{C_A}{2}\right) \left[ \ln\left(\frac{s-m_t^2}{m_t \sqrt{s}}\right)+2\ln\left(\frac{t(t-m_t^2)}{u(u-m_t^2)}\right)\right]
+\frac{C_A}{2}\ln\left(\frac{t(t-m_t^2)}{m_t s^{3/2}}\right)-\frac{C_F}{2} \, ,
\nonumber \\
\label{g1l}
\eeqa
where $C_F=(N_c^2-1)/(2N_c)$ and $C_A=N_c$, with $N_c=3$ the number of colors.

At two loops, the four matrix elements are given by \cite{NKnnll,NK3l}
\beqa
\Gamma_{S \, 11}^{(2)} &=& K_2 \, \Gamma_{S \, 11}^{(1)}+\frac{1}{4} C_F C_A (1-\zeta_3) \, ,
\nonumber \\
\Gamma_{S \, 12}^{(2)} &=& K_2 \, \Gamma_{S \, 12}^{(1)} \, , \quad \quad
\Gamma_{S \, 21}^{(2)} = K_2 \, \Gamma_{S \, 21}^{(1)} \, , 
\nonumber \\
\Gamma_{S \, 22}^{(2)} &=& K_2 \, \Gamma_{S \, 22}^{(1)}+\frac{1}{4} C_F C_A (1-\zeta_3) \, , 
\label{g2l}
\eeqa
where $K_2=C_A(67/36-\zeta_2/2)-5n_f/18$ with $n_f$ the number of light-quark flavors. 

\subsection{Soft anomalous dimension for massive final-state $b$-quarks}

When the final-state $b$-quarks are treated as massive, the expressions for the soft anomalous dimension get more complicated. I present new explicit analytical results for this case using the same color basis as in the previous subsection.

We define
\beq
L_{\beta_t \beta_b}=\frac{(1+\beta_t \beta_b)}{2(\beta_t +\beta_b)}\ln\left(\frac{(1-\beta_t)(1-\beta_b)}{(1+\beta_t)(1+\beta_b)}\right)
\eeq
where $\beta_t=\left(1-\frac{4m_t^2 s}{(s+m_t^2-m_b^2)^2}\right)^{1/2}$ 
and
$\beta_b=\left(1-\frac{4m_b^2 s}{(s+m_b^2-m_t^2)^2}\right)^{1/2}$. 

Then, the four matrix elements at one loop are given by
\beqa
\Gamma_{S\, 11}^{(1)}&=&\Gamma_{\rm cusp}^{\beta_t \beta_b \, (1)}=C_F \left[-L_{\beta_t \beta_b}-1\right] \, ,
\nonumber \\
\Gamma_{S\, 12}^{(1)}&=&\frac{C_F}{2N_c} \ln\left(\frac{(t-m_t^2)(t-m_b^2)}{(u-m_t^2)(u-m_b^2)}\right) \, , \quad \quad
\Gamma_{S\, 21}^{(1)}= \ln\left(\frac{(t-m_t^2)(t-m_b^2)}{(u-m_t^2)(u-m_b^2)}\right) \, ,
\nonumber \\
\Gamma_{S\, 22}^{(1)}&=& \!\!
\left(C_F-\frac{C_A}{2}\right) \left[-L_{\beta_t \beta_b}+2\ln\left(\frac{(t-m_t^2)(t-m_b^2)}{(u-m_t^2)(u-m_b^2)}\right)\right]
+\frac{C_A}{2} \ln\left(\frac{(t-m_t^2)(t-m_b^2)}{s \, m_t \, m_b}\right)-C_F .
\nonumber \\ 
\label{g1lm}
\eeqa

The first matrix element in Eq. (\ref{g1lm}) is the one-loop massive cusp anomalous dimension \cite{NK2loop,NK4loop} but with different masses for the two final-state lines. We note that in the massless limit, $m_b \rightarrow 0$, we have $L_{\beta_t \beta_b} \rightarrow -\ln((s-m_t^2)/(m_t m_b))$.
Then, the massive $b$-quark result for $\Gamma_S$ in Eq. (\ref{g1lm}) reduces to the massless $b$-quark result in Eq. (\ref{g1l}) if, in addition, we account for self-energies by adding the term $(C_F/2)[\ln(m_b^2/s)+1]$ in the diagonal matrix elements of Eq. (\ref{g1lm}). This provides a nice consistency check of the massive result.

We also note that if, on the other hand, we set $m_b$ equal to $m_t$, then Eq. (\ref{g1lm}) reproduces the one-loop soft anomalous dimension matrix for top-pair production via $q{\bar q} \rightarrow t {\bar t}$ \cite{NKGS1,NKGS2}.

At two loops, the four matrix elements are given by
\beqa
&& \hspace{-8mm} \Gamma^{(2)}_{S\, 11}=\Gamma_{\rm cusp}^{\beta_t \beta_b \, (2)}, \, \hspace{2mm}
\Gamma^{(2)}_{12}=
\left(K_2-C_A \, N_2^{\beta_t \beta_b}\right) \Gamma^{(1)}_{12}, \, \hspace{2mm}
\Gamma^{(2)}_{21}=
\left(K_2+C_A \, N_2^{\beta_t \beta_b}\right) \Gamma^{(1)}_{21} \, ,
\nonumber \\ && \hspace{-8mm}
\Gamma^{(2)}_{22}=
K_2 \Gamma^{(1)}_{22}
+\left(1-\frac{C_A}{2C_F}\right)
\left(\Gamma_{\rm cusp}^{\beta_t \beta_b \, (2)}-K_2 \Gamma_{\rm cusp}^{\beta_t \beta_b \, (1)}\right)
+\frac{1}{4} C_A^2(1-\zeta_3) \, .
\label{g2lm}
\eeqa

The first matrix element in Eq. (\ref{g2lm}) is the two-loop massive cusp anomalous dimension \cite{NK2loop,NK4loop} but with different masses for the two final-state lines, and its explicit expression is
\beqa
\Gamma^{\beta_t \beta_b \, (2)}_{\rm cusp}&=&K_2 \, \Gamma^{\beta_t \beta_b\, (1)}_{\rm cusp}+C_F C_A \left\{\frac{1}{2}+\frac{\zeta_2}{2} \right.
+\frac{1}{8}\ln^2\left(\frac{(1-\beta_t)(1-\beta_b)}{(1+\beta_t)(1+\beta_b)}\right) 
\nonumber \\ && 
{}+\frac{(1+\beta_t \beta_b)}{2(\beta_t+\beta_b)}\left[\frac{\zeta_2}{2}\ln\left(\frac{(1-\beta_t)(1-\beta_b)}{(1+\beta_t)(1+\beta_b)}\right)-\frac{1}{4}\ln^2\left(\frac{(1-\beta_t)(1-\beta_b)}{(1+\beta_t)(1+\beta_b)}\right) \right.
\nonumber \\ && \hspace{25mm}\left.
{}+\frac{1}{24}\ln^3\left(\frac{(1-\beta_t)(1-\beta_b)}{(1+\beta_t)(1+\beta_b)}\right)
-{\rm Li}_2\left(\frac{2(\beta_t+\beta_b)}{(1+\beta_t)(1+\beta_b)}\right)\right] 
\nonumber \\ && 
{}+\frac{(1+\beta_t\beta_b)^2}{2(\beta_t+\beta_b)^2}\left[-\zeta_3-\frac{\zeta_2}{2}\ln\left(\frac{(1-\beta_t)(1-\beta_b)}{(1+\beta_t)(1+\beta_b)}\right)-\frac{1}{24}\ln^3\left(\frac{(1-\beta_t)(1-\beta_b)}{(1+\beta_t)(1+\beta_b)}\right) \right.
\nonumber \\ &&  \hspace{5mm} \left. \left.
{}-\frac{1}{2}\ln\left(\frac{(1-\beta_t)(1-\beta_b)}{(1+\beta_t)(1+\beta_b)}\right) {\rm Li}_2\left(\frac{(1-\beta_t)(1-\beta_b)}{(1+\beta_t)(1-\beta_b)}\right) +{\rm Li}_3\left(\frac{(1-\beta_t)(1-\beta_b)}{(1+\beta_t)(1+\beta_b)}\right)\right] \right\} \, .
\nonumber \\
\label{2loopcusp}
\eeqa

Also, we have
\beqa
N_2^{\beta_t \beta_b}&=&\frac{1}{16}\ln^2\left(\frac{(1-\beta_t)(1-\beta_b)}{(1+\beta_t)(1+\beta_b)}\right)
+\frac{(1+\beta_t \beta_b)}{4(\beta_t+\beta_b)} \left[\zeta_2
-\frac{1}{4}\ln^2\left(\frac{(1-\beta_t)(1-\beta_b)}{(1+\beta_t)(1+\beta_b)}\right) \right.
\nonumber \\ && \hspace{20mm} \left.
-{\rm Li}_2\left(\frac{2(\beta_t+\beta_b)}{(1+\beta_t)(1+\beta_b)}\right)\right] \, .
\eeqa

We note that the massive $b$-quark two-loop result in Eq. (\ref{g2lm}) reduces to the massless $b$-quark two-loop result in Eq. (\ref{g2l}) in the limit $m_b \to 0$ if we also account for self-energies by adding the terms $(K_2/2)C_F-C_F C_A (1-\zeta_3)/4+(C_F/2)K_2 \ln(m_b^2/s)$ in the diagonal matrix elements of Eq. (\ref{g2lm}). This, again, provides a nice consistency check of the massive result.

We also note that if, on the other hand, we set $m_b$ equal to $m_t$, then Eq. (\ref{g2lm}) reproduces the two-loop soft anomalous dimension matrix for top-pair production via $q{\bar q} \rightarrow t {\bar t}$ \cite{NKtt2l}.

\mysection{Cross sections at LHC energies}

In this section we provide numerical results for single-top and single-antitop $s$-channel production cross sections at LHC energies treating the final-state $b$-quark as massless. We first compare with the results in Ref. \cite{NNLO} using CT14 NNLO pdf \cite{CT14} and $m_t=172.5$ GeV and other parameters as stated in Ref. \cite{NNLO}.
 
We start our comparison with Ref. \cite{NNLO} for single-top $s$-channel production. At 8 TeV energy, the NLO corrections provide a 35.4\% increase over the LO result, with NNLO giving a further 7.4\% increase, for a total of 43\% higher-order corrections \cite{NNLO}. We confirm the LO result and the NLO result, and if we define as approximate NNLO (aNNLO) the sum of the NLO cross section and the second-order soft-gluon corrections, we find that our aNNLO result of 3.54 pb with central scale $\mu=m_t$ is within around 7 per mille of the NNLO cross section given in Ref. \cite{NNLO}. Furthermore, the scale variation of our aNNLO calculation over $m_t/2 \le \mu \le 2m_t$ is small and very similar to that in \cite{NNLO}, i.e. 2\%. This shows that the soft-gluon corrections are clearly dominant and that the soft-gluon approximation is excellent. At 13 TeV energy, the NLO corrections provide a 35.0\% increase over the LO result, with NNLO giving a further 6.9\% increase, for a total of 42\% higher-order corrections \cite{NNLO}. Again, we agree with the LO and NLO results and find that our central aNNLO result of 6.74 pb is within around 5 per mille of the NNLO cross section in Ref. \cite{NNLO}, with a similar small scale uncertainty of 1\%, showing that the approximation works very well.

We continue our comparison with \cite{NNLO} for single-antitop $s$-channel production. At 8 TeV energy, the NLO corrections provide a 35.5 \% increase over the LO result, with NNLO giving a further 7.5\%, for a total of 43\% corrections. We confirm the LO and NLO results, and find that our central aNNLO result of 2.01 pb is within around 7 per mille of the NNLO cross section in \cite{NNLO} and with a similar scale uncertainty of 2\%. At 13 TeV energy, the NLO corrections provide a 34.9\% increase over the LO result, with NNLO giving a further 6.9\%, for a total of 42\% corrections. Again, we agree with the LO and NLO results and find that our aNNLO result of 4.23 pb is within around 4 per mille of the NNLO cross section in \cite{NNLO} with a similar scale uncertainty of 1\%. Again, these results confirm the excellence of the soft-gluon approximation.

Next, we give updated predictions using MSHT20 aN$^3$LO pdf \cite{MSHT20an3lo}, $m_t=172.5$ GeV, and the latest values for other parameters as given in the 2024 Review of Particle Physics \cite{RPP24}. In addition, we provide the aN$^3$LO cross sections by adding third-order soft-gluon corrections to the NNLO results.

\begin{table}[htb]
\begin{center}
\begin{tabular}{|c|c|c|c|c|c|} \hline
\multicolumn{6}{|c|}{Single-top $s$-channel cross sections in $pp$ collisions} \\ \hline
$\sigma$ in pb & 7 TeV & 8 TeV & 13 TeV & 13.6 TeV & 14 TeV \\ \hline
LO   & $2.11^{+0.01}_{-0.01}$ & $2.55^{+0.01}_{-0.01}$ & $4.85^{+0.12}_{-0.16}$ & $5.14^{+0.14}_{-0.18}$ & $5.33^{+0.15}_{-0.20}$ \\ \hline
NLO  & $2.85^{+0.07}_{-0.05}$ & $3.46^{+0.08}_{-0.06}$ & $6.57^{+0.09}_{-0.06}$ & $6.98^{+0.10}_{-0.06}$ & $7.25^{+0.10}_{-0.07}$ \\ \hline
NNLO & $3.01^{+0.03}_{-0.03}$ & $3.65^{+0.03}_{-0.03}$ & $6.91^{+0.05}_{-0.04}$ & $7.34^{+0.06}_{-0.04}$ & $7.62^{+0.06}_{-0.04}$ \\ \hline
aN$^3$LO & $3.09^{+0.02}_{-0.03}$ & $3.74^{+0.02}_{-0.03}$ & $7.07^{+0.03}_{-0.04}$ & $7.51^{+0.04}_{-0.04}$ & $7.79^{+0.05}_{-0.04}$  \\ \hline
\end{tabular}
\caption[]{The single-top $s$-channel cross sections (in pb) at different perturbative orders in $pp$ collisions with various values of $\sqrt{S}$, with $m_t=172.5$ GeV and MSHT20 aN$^3$LO pdf. The central results are with $\mu=m_t$, and the uncertainties are from scale variation over $m_t/2 \le \mu \le 2m_t$.}
\label{table1}
\end{center}
\end{table}

In Table \ref{table1}, we present the single-top $s$-channel cross sections at LO, NLO, NNLO, and aN$^3$LO in proton-proton collisions for LHC energies of 7, 8, 13, 13.6, and 14 TeV. The NNLO results presented here are calculated by multiplying the results in \cite{NNLO} by $K$-factors to account for the different pdf and parameters (these $K$-factors are simply the ratios of the aNNLO results with the new pdf/parameters divided by those with the old ones). The central results are with the scale choice $\mu=m_t$ and the scale uncertainties indicated are from variation over the interval $m_t/2 \le \mu \le 2m_t$. Including both scale and pdf uncertainties, the aN$^3$LO $s$-channel single-top cross section is $3.09^{+0.02}_{-0.03}{}^{+0.06}_{-0.07}$ pb at 7 TeV, $3.74^{+0.02}_{-0.03}{}^{+0.06}_{-0.09}$ pb at 8 TeV, $7.07^{+0.03}_{-0.04}{}^{+0.09}_{-0.17}$ pb at 13 TeV, $7.51^{+0.04}_{-0.04}{}^{+0.10}_{-0.18}$ pb at 13.6 TeV, and $7.79^{+0.05}_{-0.04}{}^{+0.10}_{-0.18}$ pb at 14 TeV.

\begin{table}[htb]
\begin{center}
\begin{tabular}{|c|c|c|c|c|c|} \hline
\multicolumn{6}{|c|}{Single-antitop $s$-channel cross sections in $pp$ collisions} \\ \hline
$\sigma$ in pb & 7 TeV & 8 TeV & 13 TeV & 13.6 TeV & 14 TeV \\ \hline
LO   & $1.16^{+0.00}_{-0.01}$ & $1.45^{+0.00}_{-0.01}$ & $3.03^{+0.08}_{-0.11}$ & $3.23^{+0.10}_{-0.12}$ & $3.37^{+0.10}_{-0.13}$ \\ \hline
NLO  & $1.57^{+0.04}_{-0.03}$ & $1.97^{+0.05}_{-0.04}$ & $4.07^{+0.05}_{-0.04}$ & $4.35^{+0.06}_{-0.04}$ & $4.50^{+0.06}_{-0.04}$  \\ \hline
NNLO  & $1.66^{+0.02}_{-0.02}$ & $2.07^{+0.02}_{-0.02}$ & $4.28^{+0.03}_{-0.02}$ & $4.58^{+0.03}_{-0.02}$ & $4.74^{+0.03}_{-0.02}$  \\ \hline 
aN$^3$LO  & $1.70^{+0.02}_{-0.02}$ & $2.12^{+0.02}_{-0.02}$ & $4.38^{+0.02}_{-0.02}$ & $4.69^{+0.02}_{-0.02}$ & $4.85^{+0.02}_{-0.02}$  \\ \hline
\end{tabular}
\caption[]{The single-antitop $s$-channel cross sections (in pb) at different perturbative orders in $pp$ collisions with various values of $\sqrt{S}$, with $m_t=172.5$ GeV and MSHT20 aN$^3$LO pdf. The central results are with $\mu=m_t$, and the uncertainties are from scale variation over $m_t/2 \le \mu \le 2m_t$.}
\label{table2}
\end{center}
\end{table}

In Table \ref{table2}, we present the single-antitop $s$-channel cross sections at various orders for LHC energies of 7, 8, 13, 13.6, and 14 TeV. Again, the central results are with the scale choice $\mu=m_t$ and the scale variation is over the interval $m_t/2 \le \mu \le 2m_t$. Including both scale and pdf uncertainties, the aN$^3$LO $s$-channel single-antitop cross section is $1.70^{+0.02}_{-0.02}{}^{+0.04}_{-0.04}$ pb at 7 TeV, $2.12^{+0.02}_{-0.02}{}^{+0.04}_{-0.05}$ pb at 8 TeV, $4.38^{+0.02}_{-0.02}{}^{+0.04}_{-0.12}$ pb at 13 TeV, $4.69^{+0.02}_{-0.02}{}^{+0.05}_{-0.12}$ pb at 13.6 TeV, and $4.85^{+0.02}_{-0.02}{}^{+0.05}_{-0.13}$ pb at 14 TeV.

\begin{table}[!htb]
\begin{center}
\begin{tabular}{|c|c|c|c|c|c|} \hline
\multicolumn{6}{|c|}{Single-top plus single-antitop $s$-channel cross sections in $pp$ collisions} \\ \hline
$\sigma$ in pb & 7 TeV & 8 TeV & 13 TeV & 13.6 TeV & 14 TeV \\ \hline
LO   & $3.27^{+0.01}_{-0.02}$ & $4.00^{+0.01}_{-0.02}$ & $7.88^{+0.20}_{-0.27}$ & $8.37^{+0.24}_{-0.30}$ & $8.70^{+0.25}_{-0.33}$ \\ \hline
NLO  & $4.42^{+0.11}_{-0.08}$ & $5.43^{+0.13}_{-0.10}$ & $10.6^{+0.1}_{-0.1}$ & $11.3^{+0.2}_{-0.1}$ & $11.8^{+0.2}_{-0.1}$ \\ \hline
NNLO  & $4.67^{+0.05}_{-0.04}$ & $5.72^{+0.06}_{-0.05}$ & $11.2^{+0.08}_{-0.06}$ & $11.9^{+0.09}_{-0.06}$ & $12.4^{+0.09}_{-0.06}$  \\ \hline 
aN$^3$LO  & $4.79^{+0.04}_{-0.04}$ & $5.86^{+0.04}_{-0.05}$ & $11.5^{+0.05}_{-0.06}$ & $12.2^{+0.06}_{-0.06}$ & $12.6^{+0.07}_{-0.06}$  \\ \hline
\end{tabular}
\caption[]{The single-top plus single-antitop $s$-channel cross sections (in pb) at different perturbative orders in $pp$ collisions with various values of $\sqrt{S}$, with $m_t=172.5$ GeV and MSHT20 aN$^3$LO pdf. The central results are with $\mu=m_t$, and the uncertainties are from scale variation over $m_t/2 \le \mu \le 2m_t$.}
\label{table3}
\end{center}
\end{table}

Table \ref{table3} shows the sum of the single-top and single-antitop $s$-channel cross sections through aN$^3$LO. Including both scale and pdf uncertainties, the aN$^3$LO $s$-channel single-top plus single-antitop cross section is $4.79^{+0.04}_{-0.04}{}^{+0.10}_{-0.11}$ pb at 7 TeV, $5.86^{+0.04}_{-0.05}{}^{+0.11}_{-0.14}$ pb at 8 TeV, $11.5^{+0.05}_{-0.06}{}^{+0.13}_{-0.29}$ pb at 13 TeV, $12.2^{+0.06}_{-0.06}{}^{+0.15}_{-0.30}$ pb at 13.6 TeV, and $12.6^{+0.07}_{-0.06}{}^{+0.16}_{-0.32}$ pb at 14 TeV. 

\begin{figure}[htbp]
\begin{center}
\includegraphics[width=10cm]{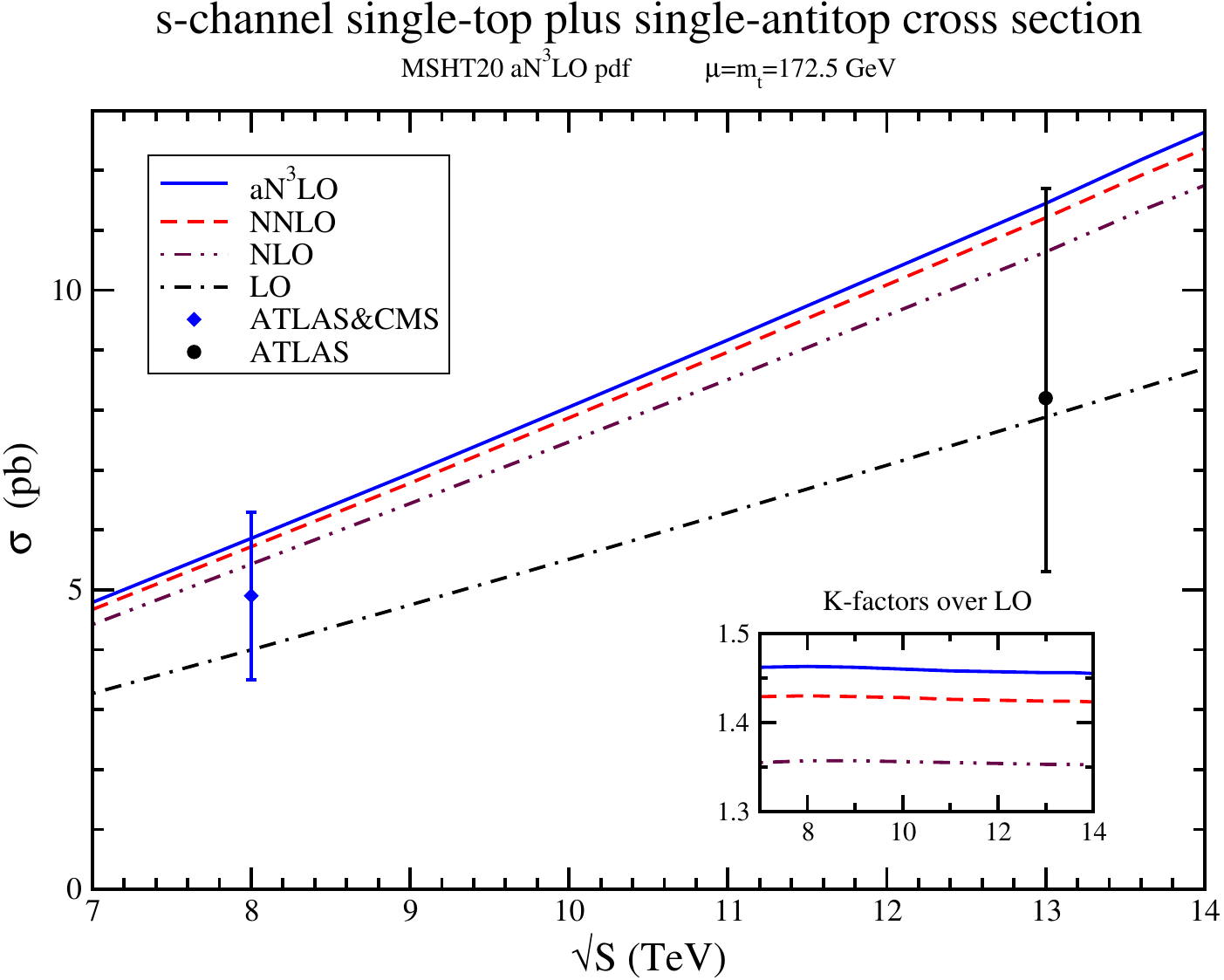}
\caption{The total cross sections for $s$-channel single-top plus single-antitop production in $pp$ collisions at LHC energies at LO, NLO, NNLO, and aN$^3$LO. Also shown for comparison are LHC data at 8 TeV \cite{ATLASCMS} and 13 TeV \cite{ATLAS13} energies. The inset plot displays the $K$-factors of the higher-order cross sections relative to LO.}
\label{schannelplot}
\end{center}
\end{figure}

In Fig. \ref{schannelplot}, we plot the central results for the total $s$-channel single-top plus single-antitop cross section as a function of LHC energy. Results are shown at LO, NLO, NNLO, and aN$^3$LO. Comparisons are made with the ATLAS+CMS combination of measured cross sections of $4.9 \pm 1.4$ pb at 8 TeV \cite{ATLASCMS}, and with the ATLAS measured cross section of $8.2^{+3.5}_{-2.9}$ pb at 13 TeV \cite{ATLAS13}. The data are in agreement with the theoretical predictions but the experimental errors bars are currently large. The inset plot in Fig. \ref{schannelplot} shows the $K$-factors, i.e. the ratios of the higher-order cross sections to the LO result. We note that the $K$-factors are fairly stable over the LHC energy range.

\mysection{Conclusions}

The production of single top and single antitop quarks via the $s$-channel processes $q{\bar q'} \to t{\bar b}$ and $q{\bar q'} \to {\bar t}b$, which involve a timelike virtual $W$ boson, has been a topic of intense study at the Tevatron and the LHC. I have shown in this paper that the soft-gluon corrections dominate the cross section through NNLO hence demonstrating once again, as for many other top-quark processes, the relevance and excellence of the soft-gluon approximation. Moreover, I have calculated third-order soft-gluon corrections and added them to the NNLO cross sections to derive aN$^3$LO predictions for the cross sections. The higher-order corrections reduce the scale dependence of the theoretical prediction. The experimental error bars are currently quite large but the data is in agreement with the theoretical calculations.

The numerical calculations assume massless $b$-quarks in the final state. I have also shown how to extend the resummation formalism to the case of massive final-state $b$-quarks, and I have provided explicit analytical results for the soft anomalous dimension matrix at one and two loops.

\mysection*{Acknowledgements}
This material is based upon work supported by the National Science Foundation under Grant No. PHY 2412071.


\begin{thebibliography}{99}

\bibitem{D02009}
D0 Collaboration,  {\sl Observation of single top-quark production}, Phys. Rev. Lett. {\bf 103}, 092001 (2009) [arXiv:0903.0850].

\bibitem{CDF2009}
CDF Collaboration, {\sl Observation of electroweak single top-quark production}, Phys. Rev. Lett. {\bf 103}, 092002 (2009) [arXiv:0903.0885].

\bibitem{CDF2010}
CDF Collaboration, {\sl Observation of single top quark production and measurement of $|V_{tb}|$ with CDF}, Phys. Rev. D {\bf 82}, 112005 (2010) [arXiv:1004.1181].

\bibitem{D02011}
D0 Collaboration, {\sl Measurements of single top quark production cross sections and $|V_{tb}|$ in $p{\bar p}$ collisions at ${\sqrt s}=1.96$ TeV}, Phys. Rev. D {\bf 84}, 112001 (2011) [arXiv:1108.3091].

\bibitem{D02013}
D0 Collaboration, {\sl Evidence for $s$-channel single top quark production in $p{\bar p}$ collisions at ${\sqrt s}=1.96$ TeV}, Phys. Lett. B {\bf 726}, 656 (2013) [arXiv:1307.0731].

\bibitem{CDF2014}
CDF Collaboration, {\sl Evidence for $s$-channel single-top-quark production in events with one charged lepton and two jets at CDF}, Phys. Rev. Lett. {\bf 112}, 231804 (2014) [arXiv:1402.0484].

\bibitem{CDF2014p}
CDF Collaboration, {\sl Search for $s$-channel single-top-quark production in events with missing energy plus jets in $p{\bar p}$ collisions at ${\sqrt s}=1.96$ TeV}, Phys. Rev. Lett. {\bf 112}, 231805 (2014) [arXiv:1402.3756].

\bibitem{CDFD02014}
CDF and D0 Collaborations, {\sl Observation of $s$-channel production of single top quarks at the Tevatron}, Phys. Rev. Lett. {\bf 112}, 231803 (2014) [arXiv:1402.5126].

\bibitem{CDF2016}
CDF Collaboration, {\sl Measurement of the single top quark production cross section and $|V_{tb}|$ in 1.96 TeV $p{\bar p}$ collisions with missing transverse energy and jets and final CDF combination}, Phys. Rev. D {\bf 93}, 032011 (2016) [arXiv:1410.4909].

\bibitem{CDFD02015}
CDF and D0 Collaborations, {\sl Tevatron combination of single-top-quark cross sections and determination of the magnitude of the Cabibbo-Kobayashi-Maskawa matrix element $|V_{tb}|$}, Phys. Rev. Lett. {\bf 115}, 152003 (2015) [arXiv:1503.05027].

\bibitem{ATLAS7}
ATLAS Collaboration, {\sl Search for $s$-channel single top-quark production in $pp$ collisions at ${\sqrt s}=7$ TeV}, ATLAS-CONF-2011-118. 

\bibitem{ATLAS8a}
ATLAS Collaboration, {\sl Search for $s$-channel single top-quark production in proton-proton collisions at ${\sqrt s}=8$ TeV with the ATLAS detector}, Phys. Lett. B {\bf 740}, 118 (2015) [arXiv:1410.0647].

\bibitem{ATLAS8b}
ATLAS Collaboration, {\sl Evidence for single top-quark production in the $s$-channel in proton-proton collisions at ${\sqrt s}=8$ TeV with the ATLAS detector using the Matrix Element Method}, Phys. Lett. B {\bf 756}, 228 (2016) [arXiv:1511.05980].

\bibitem{CMS7and8}
CMS Collaboration, {\sl Search for $s$ channel single top quark production in $pp$ collisions at ${\sqrt s}= 7$ and 8 TeV}, JHEP {\bf 09}, 027 (2016) [arXiv:1603.02555].

\bibitem{ATLASCMS}
ATLAS and CMS Collaborations, {\sl Combinations of single-top-quark production cross-section measurements and $|f_{LV} V_{tb}|$ determinations at ${\sqrt s}=7$ and 8 TeV with the ATLAS and CMS experiments}, JHEP {\bf 05}, 088 (2019) [arXiv:1902.07158].

\bibitem{ATLAS13}
ATLAS Collaboration, {\sl Measurement of single top-quark production in the $s$-channel in proton-proton collisions at ${\sqrt s}=13$ TeV with the ATLAS detector}, JHEP {\bf 06}, 191 (2023) [arXiv:2209.08990].

\bibitem{NLO1}
M.C. Smith and S.S. Willenbrock, {\sl QCD and Yukawa corrections to single top quark production via $q{\bar q} \to t {\bar b}$}, Phys. Rev. D {\bf 54}, 6696 (1996) [arXiv:hep-ph/9604223].

\bibitem{NLO2}
B.W. Harris, E. Laenen, L. Phaf, Z. Sullivan, and S. Weinzierl, {\sl Fully differential single-top-quark cross section in next-to-leading order QCD}, Phys. Rev. D {\bf 66}, 054024 (2002) [arXiv:hep-ph/0207055].

\bibitem{NLO3}
Z. Sullivan, {\sl Understanding single-top-quark production and jets at hadron colliders}, Phys. Rev. D {\bf 70}, 114012 (2004) [arXiv:hep-ph/0408049].

\bibitem{NLO4}
J. Campbell, R.K. Ellis, and F. Tramontano, {\sl Single top-quark production and decay at next-to-leading order}, Phys. Rev. D {\bf 70}, 094012 (2004) [arXiv:hep-ph/0408158].

\bibitem{NLO5}
Q.-H. Cao and C.-P. Yuan, {\sl Single top quark production and decay at next-to-leading order in hadron collisions}, Phys. Rev. D {\bf 71}, 054022 (2005) [arXiv:hep-ph/0408180].

\bibitem{NLO6}
Q.-H. Cao, R. Schwienhorst, and C.-P. Yuan, {\sl Next-to-leading order corrections to single top quark production and decay at the Fermilab Tevatron: $s$-channel process}, Phys. Rev. D {\bf 71}, 054023 (2005) [arXiv:hep-ph/0409040]. 

\bibitem{NLO7}
S. Heim, Q.-H. Cao, R. Schwienhorst, and C.-P. Yuan, {\sl Next-to-leading order QCD corrections to $s$-channel single top quark production and decay at the LHC}, Phys. Rev. D {\bf 81}, 034005 (2010) [arXiv:0911.0620].

\bibitem{NLO8}
P. Falgari, F. Giannuzzi, P. Mellor, and A. Signer, {\sl Off-shell effects for $t$-channel and $s$-channel single-top production at next-to-leading order in QCD}, Phys. Rev. D {\bf 83}, 094013 (2011) [arXiv:1102.5267].

\bibitem{NNLO}
Z.L. Liu and J. Gao, {\sl $s$-channel single top quark production and decay at next-to-next-to-leading-order in QCD}, Phys. Rev. D {\bf 98}, 071501 (2018) [arXiv:1807.03835].

\bibitem{NKnlltev}
N. Kidonakis, {\sl Single top quark production at the Fermilab Tevatron: Threshold resummation and finite-order soft gluon corrections}, Phys. Rev. D {\bf 74}, 114012 (2006) [arXiv:hep-ph/0609287].

\bibitem{NKnlllhc}
N. Kidonakis, {\sl Higher-order soft gluon corrections in single top quark production at the CERN LHC}, Phys. Rev. D {\bf 75}, 071501 (2007) [arXiv:hep-ph/0701080].

\bibitem{NKnnll}
N. Kidonakis, {\sl Next-to-next-to-leading logarithm resummation for $s$-channel single top quark production}, Phys. Rev. D {\bf 81}, 054028 (2010) [arXiv:1001.5034].

\bibitem{PPN}
N. Kidonakis, {\sl NNLL threshold resummation for top-pair and single-top production}, Phys. Part. Nucl. {\bf 45}, 714 (2014) [arXiv:1210.7813].

\bibitem{HQ13}
N. Kidonakis, {\sl Top quark production}, Physics of Heavy Quarks and Hadrons (HQ2013), p. 139 [arXiv:1311.0283].

\bibitem{NKdiff}
N. Kidonakis, {\sl Single-top transverse-momentum distributions at approximate NNLO}, Phys. Rev. D {\bf 93}, 054022 (2016) [arXiv:1510.06361].

\bibitem{NKrev}
N. Kidonakis, {\sl Soft-gluon corrections in top-quark production}, Int. J. Mod. Phys. A {\bf 33}, 1830021 (2018) [arXiv:1806.03336].

\bibitem{SYY}
P. Sun, B. Yan, and C.-P. Yuan, {\sl Transverse momentum resummation for $s$-channel single top quark production at the LHC}, Phys. Rev. D {\bf 99}, 034008 (2019) [arXiv:1811.01428].

\bibitem{NK3l}
N. Kidonakis, {\sl Soft anomalous dimensions for single-top production at three loops}, Phys. Rev. D {\bf 99}, 074024 (2019) [arXiv:1901.09928].

\bibitem{NKaN3LO}
N. Kidonakis, {\sl Next-to-next-to-next-to-leading-order soft-gluon corrections in hard-scattering processes near threshold}, Phys. Rev. D {\bf 73}, 034001 (2006) [arXiv:hep-ph/0509079]. 

\bibitem{NKtt2l}
N. Kidonakis, {\sl Next-to-next-to-leading soft-gluon corrections for the top quark cross section and transverse momentum distribution}, Phys. Rev. D {\bf 82}, 114030 (2010) [arXiv:1009.4935]. 

\bibitem{NKtW}
N. Kidonakis, {\sl Two-loop soft anomalous dimensions for single top quark associated production with a $W^-$ or $H^-$}, Phys. Rev. D {\bf 82}, 054018 (2010) [arXiv:1005.4451]. 

\bibitem{NKtW2}
N. Kidonakis, {\sl Soft-gluon corrections for $tW$ production at N$^3$LO}, Phys. Rev. D {\bf 96}, 034014 (2017) [arXiv:1612.06426].
    
\bibitem{NKtch}
N. Kidonakis, {\sl Next-to-next-to-leading-order collinear and soft gluon corrections for $t$-channel single top quark production}, Phys. Rev. D {\bf 83}, 091503 (2011) [arXiv:1103.2792]. 

\bibitem{MFNK}
M. Forslund and N. Kidonakis, {\sl Resummation for $2 \to n$ processes in single-particle-inclusive kinematics}, Phys. Rev. D {\bf 102}, 034006 (2020) [arXiv:2003.09021].

\bibitem{tqH}
M. Forslund and N. Kidonakis, {\sl Soft-gluon corrections for the associated production of a single top quark and a Higgs boson}, Phys. Rev. D {\bf 104}, 034024 (2021) [arXiv:2103.01228].

\bibitem{tqgamma} 
N. Kidonakis and N. Yamanaka, {\sl QCD corrections in $tq\gamma$ production at hadron colliders}, Eur. Phys. J. C {\bf 82}, 670 (2022) [arXiv:2201.12877].

\bibitem{tqZ} 
N. Kidonakis and N. Yamanaka, {\sl Soft-gluon corrections for $tqZ$ production}, Phys. Lett. B {\bf 838}, 137708 (2023) [arXiv:2210.09542].

\bibitem{ttgamma} 
N. Kidonakis and A. Tonero, {\sl Higher-order corrections in $t{\bar t}\gamma$ cross sections}, Phys. Rev. D {\bf 107}, 034013 (2023) [arXiv:2212.00096].

\bibitem{ttW} 
N. Kidonakis and C. Foster, {\sl Soft-gluon corrections in $t{\bar t}W$ production}, Phys. Lett. B {\bf 854}, 138708 (2024) [arXiv:2312.00861]. 

\bibitem{ttZ}
N. Kidonakis and C. Foster, {\sl Higher-order soft-gluon corrections for $t{\bar t}Z$ cross sections}, Phys. Lett. B {\bf 860}, 139146 (2025) [arXiv:2410.01214].

\bibitem{ttH}
N. Kidonakis and N. Yamanaka, {\sl Higher-order QCD and electroweak corrections for $t{\bar t}H$ production}, Phys. Lett. B {\bf 873}, 140175 (2026) [arXiv:2509.23293].

\bibitem{1PIH}
N. Kidonakis and A. Tonero, {\sl N$^3$LO soft-gluon corrections in single-particle-inclusive kinematics and $H^+ H^-$ production}, JHEP {\bf 06}, 138 (2024) [arXiv:2404.00089].

\bibitem{Wgamma}
N. Kidonakis and A. Tonero, {\sl Higher-order soft and virtual corrections in $pp \to \gamma W$ production at the LHC}, Eur. Phys. J. C {\bf 85}, 1270 (2025) [arXiv:2506.01590].

\bibitem{GS87}
G. Sterman, {\sl Summation of large corrections to short-distance hadronic cross sections}, Nucl. Phys. B {\bf 281}, 310 (1987). 

\bibitem{CT89}
S. Catani and L. Trentadue, {\sl Resummation of the QCD perturbative series for hard processes}, Nucl. Phys. B {\bf 327}, 323 (1989).

\bibitem{FNPY1}
A. Ferroglia, M. Neubert, B.D. Pecjak, and L.L. Yang, {\sl Two-loop divergences of scattering amplitudes with massive partons}, Phys. Rev. Lett. {\bf 103}, 201601 (2009) [arXiv:0907.4791].

\bibitem{FNPY2}
A. Ferroglia, M. Neubert, B.D. Pecjak, and L.L. Yang, {\sl Two-loop divergences of massive scattering amplitudes in non-abelian gauge theories}, JHEP {\bf 11}, 062 (2009) [arXiv:0908.3676].

\bibitem{ZLWZ}
H.X. Zhu, C.S. Li, J. Wang, and J.J. Zhang, {\sl Factorization and resummation of $s$-channel single top quark production}, JHEP {\bf 02}, 099 (2011) [arXiv:1006.0681].

\bibitem{NK2loop}
N. Kidonakis, {\sl Two-loop soft anomalous dimensions and next-to-next-to-leading-logarithm resummation for heavy quark production}, Phys. Rev. Lett. {\bf 102}, 232003 (2009) [arXiv:0903.2561]. 

\bibitem{NK4loop}
N. Kidonakis, {\sl Four-loop massive cusp anomalous dimension in QCD: A calculation from asymptotics}, Phys. Rev. D {\bf 107}, 054006 (2023) [arXiv:2301.05972]. 

\bibitem{NKGS1}
N. Kidonakis and G. Sterman, {\sl Subleading logarithms in QCD hard scattering}, Phys. Lett. B {\bf 387}, 867 (1996).

\bibitem{NKGS2}
N. Kidonakis and G. Sterman, {\sl Resummation for QCD hard scattering}, Nucl. Phys. B {\bf 505}, 321 (1997) [arXiv:hep-ph/9705234]. 

\bibitem{CT14}
S. Dulat, T.-J. Hou, J. Gao, M. Guzzi, J. Huston, P. Nadolsky, J. Pumplin, C. Schmidt, D. Stump, and C.-P. Yuan, {\sl New parton distribution functions from a global analysis of quantum chromodynamics}, 
Phys. Rev. D {\bf 93}, 033006 (2016) [arXiv:1506.07443].

\bibitem{MSHT20an3lo}
J. McGowan, T. Cridge, L.A. Harland-Lang, and R.S. Thorne, {\sl Approximate N$^3$LO parton distribution functions with theoretical uncertainties: MSHT20aN$^3$LO PDFs}, Eur. Phys. J. C {\bf 83}, 185 (2023) [arXiv:2207.04739].

\bibitem{RPP24}
Particle Data Group, {\sl Review of Particle Physics}, Phys. Rev. D {\bf 110}, 030001 (2024).

\end{thebibliography}
\end{document}